\documentclass[table,dvipsnames]{optica-article}
\journal{opticajournal}
\articletype{Research Article}
\usepackage{enumitem}
\usepackage{multirow}
\usepackage{hyperref}
\usepackage{upgreek}
\hypersetup{
    colorlinks=true,
    linkcolor=blue,
    citecolor=cyan,
    urlcolor=green,
}
\usepackage{xcolor,etoolbox}

\usepackage{colortbl}
\usepackage{tikz,xcolor,hyperref}
\hypersetup{colorlinks=true,linkcolor=blue,urlcolor=blue,citecolor=blue,pdfstartview={XYZ null null 1.00},pdfauthor={Tamás Csizmadia}}
\definecolor{lime}{HTML}{A6CE39}
\DeclareRobustCommand{\orcidicon}{%
	\begin{tikzpicture}
	\draw[lime, fill=lime] (0,0) 
	circle [radius=0.16] 
	node[white] {{\fontfamily{qag}\selectfont \tiny ID}};
	\draw[white, fill=white] (-0.0625,0.095) 
	circle [radius=0.007];
	\end{tikzpicture}
	\hspace{-2mm}
}

\foreach \x in {A, ..., Z}{%
	\expandafter\xdef\csname orcid\x\endcsname{\noexpand\href{https://orcid.org/\csname orcidauthor\x\endcsname}{\noexpand\orcidicon}}
}

\newcommand{\RNum}[1]{\uppercase\expandafter{\romannumeral #1\relax}}

\begin{document}
\title{Spectrally tunable ultrashort monochromatized extreme ultraviolet pulses at 100 kHz}

\author{Tam\'{a}s Csizmadia\authormark{1,*}\orcidC{}, Zolt\'{a}n Filus\authormark{1}\orcidD{}, T\'{i}mea Gr\'{o}sz\authormark{1}\orcidE{}, Peng Ye\authormark{1,2}\orcidF{}, L\'{e}n\'{a}rd Guly\'{a}s Oldal\authormark{1}\orcidG{}, Massimo De Marco\authormark{1}\orcidH{}, P\'{e}ter J\'{o}j\'{a}rt\authormark{1}\orcidI{}, Imre Seres\authormark{1}, Zsolt Bengery\authormark{1}, Barnab\'{a}s Gilicze\authormark{1}\orcidL{}, Matteo Lucchini\authormark{3,4}\orcidM{}, Mauro Nisoli\authormark{3,4}\orcidN{}, Fabio Frassetto\authormark{5}\orcidO{}, Fabio Samparisi\authormark{5}, Luca Poletto\authormark{5}\orcidQ{}, Katalin Varj\'{u}\authormark{1,6}\orcidR{}, Subhendu Kahaly\authormark{1,7}\orcidA{}, and Bal\'{a}zs Major \authormark{1}\orcidB{}}

\address{\authormark{1}ELI ALPS, ELI-HU Non-Profit Ltd., Wolfgang Sandner utca 3, H-6728 Szeged, Hungary\\
\authormark{2}Current address: Universit\'{e} Paris-Saclay, CEA, CNRS, LIDYL, 91191 Gif-sur-Yvette, France\\
\authormark{3}Institute for Photonics and Nanotechnologies, IFN-CNR, 20133 Milano, Italy\\
\authormark{4}Department of Physics, Politecnico di Milano, 20133 Milano, Italy\\
\authormark{5}Institute for Photonics and Nanotechnologies, IFN-CNR, via Trasea 7, 35131 Padova, Italy\\
\authormark{6}Department of Optics and Quantum Electronics, University of Szeged, D\'{o}m t\'{e}r 9, H-6720 Szeged, Hungary\\
\authormark{7}Institute of Physics, University of Szeged, D\'{o}m t\'{e}r 9, H-6720 Szeged, Hungary\\}

\email{\authormark{*}tamas.csizmadia@eli-alps.hu}

\begin{abstract}
We present the experimental realization of spectrally tunable, ultrashort, quasi-monochromatic extreme ultraviolet (XUV) pulses generated at 100 kHz repetition rate in a user-oriented gas high harmonic generation (GHHG) beamline of the Extreme Light Infrastructure - Attosecond Light Pulse Source (ELI ALPS) facility. Versatile spectral and temporal shaping of the XUV pulses are accomplished with a double-grating, time-delay compensated monochromator accommodating the two composing stages in a novel, asymmetrical geometry. This configuration supports the achievement of high monochromatic XUV flux ($2.8\pm0.9\times10^{10}$ photons/s at 39.7~eV selected with 700~meV FWHM bandwidth) combined with ultrashort pulse duration (4.0$\pm$0.2~fs using 12.1$\pm$0.6~fs driving pulses) and small spot size (sub-100~$\upmu$m). Focusability, spectral bandwidth, and overall photon flux of the produced radiation were investigated covering a wide range of instrumental configurations. Moreover, complete temporal (intensity and phase) characterization of the few-femtosecond monochromatic XUV pulses --- a goal that is difficult to achieve by conventional reconstruction techniques --- has been realized using ptychographic algorithm on experimentally recorded XUV-IR pump-probe traces. The presented results contribute to in-situ, time-resolved experiments accessing direct information on the electronic structure dynamics of novel target materials.
\end{abstract}

\section{Introduction}\label{sect_introduction}
 Coherent extreme ultraviolet (XUV) and soft X-ray photon sources are the primary drivers behind the modern scientific investigation of matter at spatial and temporal scales relevant to resolve their electronic structure and dynamics. Such radiation can be provided by a wide range of instruments\cite{Couprie2014}, such as X-ray lasers\cite{Rocca1999}, synchrotron light sources \cite{Couprie2008}, or free-electron lasers (FELs) \cite{Hartmann2016,Maroju2020}. The~discovery of high-order harmonic generation (HHG) through the nonlinear interaction \cite{Nayak2019,Amini2019} of an intense infrared (IR) pulse with solid or gaseous targets \cite{Krausz2009,Chatziathanasiou2017} has opened up the way to tabletop XUV light sources with various favorable traits. Radiation produced by HHG can be tuned up to several keV photon energies \cite{Chang2019,Chen2010,Seres2006,Gao2022}, it shows good spatial and temporal coherence properties, excellent beam quality, and ultrashort pulse duration down to the sub-100 attosecond regime \cite{Chang2019, Gaumnitz2017, Kuehn2017}. These features allowed for pioneering research applications enabling, among other benefits, attosecond metrology \cite{Ramasesha2016,Pfeifer2008}, femtosecond spectroscopy \cite{Geneaux2019}, high resolution nondestructive dynamic imaging of nanosystems \cite{Gardner2017,Tanksalvala2021,Brooks2022,Eschen2022}, free-electron laser seeding \cite{McNeil2007} and boosted the development of nonlinear optics in the XUV spectral range \cite{Orfanos2020,Makos2020}.\par
 Generally, the high-order harmonic spectrum is composed of a series of peaks appearing at the odd multiples of the central laser frequency. The characteristic spectral shape consists of a rapid decline in the intensity of the peaks at the first group of harmonic orders followed by a plateau region, and a dramatic drop-down in the cut-off domain. The precise manipulation of the harmonic spectral features is of utmost significance in expanding the landscape of scientific applications of such sources. In particular, the extraction of a single harmonic peak (or a part of it) from a broad high harmonic spectrum coupled with the ability to tune such a selection over a desired wavelength range opens up the possibility to probe the electronic band structure of complex materials. In~addition, transient phases can also be studied when such a monochromatic XUV source is combined with ultrashort laser pulses in a pump-probe excitation scheme. The combination of ultrafast monochromatic excitation and angle-resolved photoemission spectroscopy (ARPES) as a diagnostic technique \cite{Mathias2007,Puppin2019,Sie2019,Lee2020,Keunecke2020} provide new opportunities: their joint capabilities expand the high-resolution energy and momentum information about the solid microworld with femtosecond time-resolution providing direct access to the underlying dynamical processes. The research interests include, for example, ultrafast changes in the population of energy levels including cooling of excited carriers via electron-phonon coupling \cite{Johannsen2013}, temporal occupation of empty states of the band structure \cite{Nicholson2019}, collective excitation dynamics of phonons \cite{Schmitt2011}, observation of scattering channels and associated excited states \cite{Cacho2012}, nonequilibrium processes in correlated systems \cite{Mathias2016}, ultrafast dynamics of excitons \cite{Garratt2022}, or metal-to-insulator transitions \cite{Nicholson2018,Lee2019}.\par
 In this work, we present the detailed characterization of ultrashort, spectrally tunable monochromatized XUV pulses produced via gas HHG at 100 kHz repetition rate, and shaped by a double-grating XUV monochromator. For~the first time to our knowledge, the ultrashort monochromatic pulses were realized with a time-delay compensating monochromator that had an asymmetric arrangement of the two composing stages. This configuration yielded ultrashort monochromatic pulses with high XUV flux via HHG driven by a high average power laser source. The annularly shaped generating IR beam was filtered out from the XUV radiation using spatial separation \cite{Ye2020,Ye2022}. This procedure needed a long first stage input arm, to which a shorter second compressor stage was coupled that enabled tight focusing conditions before the target region, as well as the reduction of the required laboratory space. The paper is divided into the following sections: Section \ref{sect_generation} discusses in detail the optical layout for XUV generation and the subsequent beam shaping, including the alignment of the driving laser beam, spatial IR filtering and time-preserving monochromatization in an asymmetric geometry. Section \ref{sect_diagnostics} exhibits the comprehensive --- spatial, spectral, and temporal --- characterization of the generated radiation, supplemented with the measurement of the efficiency and overall photon flux. Finally, the conclusive remarks are summarized in Section \ref{sect_conclusion}.
\section{Generation of ultrashort monochromatized XUV pulses}\label{sect_generation}
  \begin{figure*}[ht]
  \centering
  \includegraphics[trim=25cm 36cm 33cm 7cm,clip,width=1.0\linewidth]{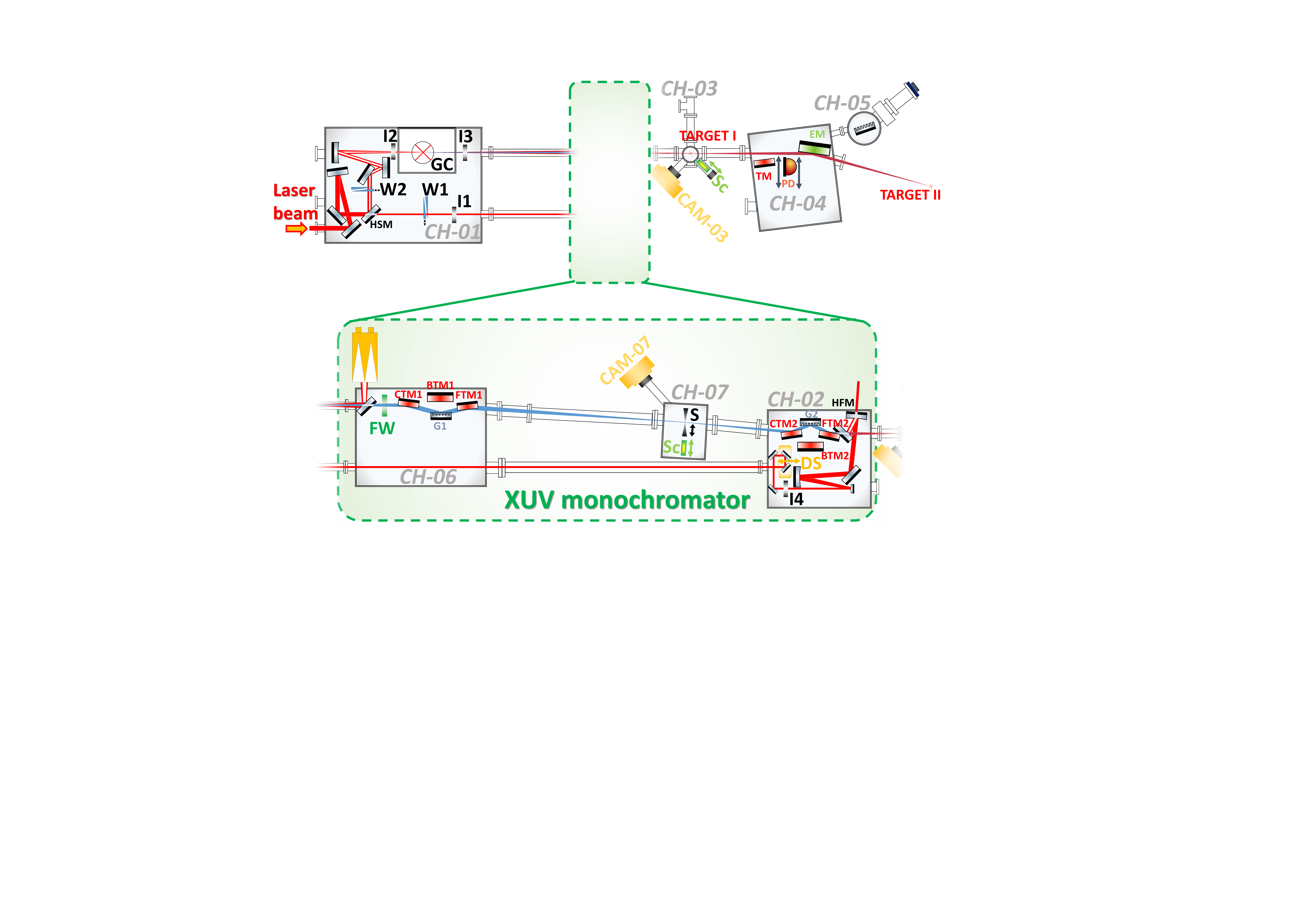}
  \caption{Schematics of the \textsc{\textit{HR GHHG Condensed}} beamline with the XUV monochromator included (circumscribed with a dashed green line). HSM: holey splitting mirror; W: wedge pair; I: iris; GC: gas cell; FW: filter wheel; CTM: collimating toroidal mirror; G: grating; FTM: focusing toroidal mirror; BTM: broadband toroidal mirror; CAM: camera; S: slit; Sc: scintillator crystal; HFM: holey focusing mirror; DS: delay stage; TM: toroidal mirror; EM: ellipsoidal mirror; PD: XUV photodiode.}
  \label{Fig_HR-GHHG_Cond}
\end{figure*}
\subsection{High-order harmonic generation at high repetition rate}
The experiments were conducted at the high repetition rate s high-order harmonic generation beamline constructed for measurements with condensed targets (\textsc{\textit{HR GHHG Condensed}}) at the ELI ALPS facility \cite{Kuehn2017,Charalambidis2017}. The beamline is driven by the HR-1 laser system (developed by Active Fiber Systems GmbH) with pulses of 1~mJ, down to 6~fs duration at 100 kHz repetition rate \cite{Hadrich2022}. It is designed to provide XUV light with photon energies ranging from 17~eV to 90~eV. The setup supports the potential of broadband spectral tuning in the generation process itself \cite{Gulyas2020,Gulyas2021,Schuster2021}. During the experiments, argon was used as target gas and phase matching conditions were fine tuned to maximize the XUV flux around the photon energy of interest. An important part of this process was the utilization of a water-cooled, custom-designed target system \cite{Filus2022} that provided the possibility to shift between various cell lengths (4, 7 or 10 mm), and to adjust the cell position along the beam propagation direction without breaking the vacuum environment. In addition, the laser output power and the backing pressure (usually a few tens of mbar) were also optimized. Upon entering the generation chamber (CH-01), the IR beam having a central wavelength of 1030~nm is directed by steering mirrors toward a holey splitting mirror (HSM), where it is divided into an annular generation and a central dressing beam (Fig.~\ref{Fig_HR-GHHG_Cond}). In both the annular and central arms, a pair of anti-reflective coated fused silica wedges (W1 and W2) are used to fine tune the dispersion of the laser pulses independently. The annularly shaped generation beam having a pulse energy of approximately 590-$\mathrm{\upmu J}$ is focused into the gas cell with a focusing mirror (focal length: 900~mm) into an estimated maximum IR intensity of about 2.7$\times10^{14}~\mathrm{Wcm^{-2}}$ and an approximately $145~\mathrm{\upmu m}$ diameter IR spot at the XUV generation point. A direct beam control and referencing system, consisting of motorized mirror mounts and optical references implemented inside the vacuum chambers, is utilized for precise beam alignment, which is critical due to the long beam path ($\approx$17~m) and high average laser power involved. Upon entering the generation chamber (CH-01), the beam is centered onto the HSM at an incident angle of 45 degrees, based on the optical image of the mirror surface and a motorized iris in the beam path of the central beam (I1), which are used as the first and second optical reference points, respectively. The annular portion of the beam is then aligned on the irises I2 and I3 in CH-01 before the XUV monochromator. The central beam passes through the chamber hosting the first monochromator stage (CH-06) without alteration, and is subsequently aligned in CH-02 with the help of another iris (I4) and the image of the hole at the center of the holey focusing mirror (HFM) before it is recombined with the XUV light for pump-probe measurements. The HFM focuses the IR light into the interaction zone (target \RNum{1}) of the first experimental chamber (CH-03), where a time-of-flight electron spectrometer (TOF, type Stefan Kaesdorf ETF11) is hosted. A spatial overlap between the XUV and IR pulses is established by the motorized tip-tilt adjustments of the HFM and the succeeding recombination mirror in CH-02, while the temporal overlap can be set with sub-7~as resolution using a delay stage (DS) implemented in the same chamber. CH-04 and 05 contain the diagnostic equipment monitoring the pulse energy and the spectral characteristics of the XUV radiation. A retractable toroidal mirror (TM) can steer the beam toward an XUV spectrometer in CH-05, which is composed of a curved, variable-line-spaced diffraction grating (HITACHI 001-0437), a microchannel plate (MCP, type Photek VID140) and a phosphor screen (P43). A CMOS camera collects the light emitted from the phosphor screen behind the MCP. The~XUV pulse energy can be measured directly after the TOF target region by a retractable XUV photodiode (PD, type NIST 40790C), which consists of a fused silica disk with a thin film (about 150 nm) of aluminum oxide deposited on top. It is absolutely calibrated in the 5--120 nm wavelength range, and is totally blind to IR radiation. Finally, if neither the XUV spectrometer nor the photodiode are in use, the beam is focused by an ellipsoidal mirror (EM) into the second target region (target \RNum{2}), where an optional end station (currently a spin- and energy-filtering photoemission microscope --- NanoESCA \cite{Wiemann2011}) is installed.\par
  \begin{table}[ht]
    \centering
    \begin{tabular}{|c|c|} 
    \hline
    \rowcolor[gray]{.9}[0.80\tabcolsep]
    \textbf{1st stage (in CH-06)} & \textbf{2nd stage (in CH-02)} \\ [1ex] 
    \hline\hline
    \rowcolor[gray]{.95}[0.60\tabcolsep]
    \multicolumn{2}{|c|}{\textbf{Low energy resolution option (100--2000~meV)}} \\
    \hline
    \textcolor{BrickRed}{G1-A: 150 gr/mm, 17-50 eV (35~eV)} & \textcolor{BrickRed}{G2-A: 300 gr/mm, 17-50 eV (31~eV)}\\
    \rowcolor[gray]{.95}[0.60\tabcolsep]
    G1-B: 300 gr/mm, 42-90 eV (60~eV) & G2-B: 600 gr/mm, 42-90 eV (52~eV)\\ [1ex]
    \hline\hline
    \multicolumn{2}{|c|}{\textbf{High energy resolution option (50--800~meV)}} \\
    \hline 
    \rowcolor[gray]{.95}[0.60\tabcolsep]
    \textcolor{BrickRed}{G1-C: 300 gr/mm, 17-42 eV (24~eV)} & \textcolor{BrickRed}{G2-C: 600 gr/mm, 17-42 eV (26~eV)} \\
    G1-D: 600 gr/mm, 25-60 eV (43~eV) & G2-D: 1200 gr/mm, 25-60 eV (43~eV)\\
    \rowcolor[gray]{.95}[0.60\tabcolsep]
    G1-E: 1200 gr/mm, 45-90 eV (70~eV) & G2-E: 2400 gr/mm, 45-90 eV (85~eV)\\
    \hline
\end{tabular}
\caption{Specifications of various optical gratings utilized in the monochromator. The~parameters of the two grid pairs used for the experimental demonstration of the temporal stretching restoration are highlighted in red. The blaze photon energy of each grating is indicated in parentheses. In the grating designation, the numbers "1" or "2" refers to the stage, in which the grating is located, while capital letters "A"--"E" mark different optimal photon energy ranges for the given optics.}
\label{tab_parameters}
\end{table}
\subsection{Spectral selection with time-delay compensation}
  \begin{figure*}[ht]
  \centering
  \includegraphics[width=1.0\linewidth]{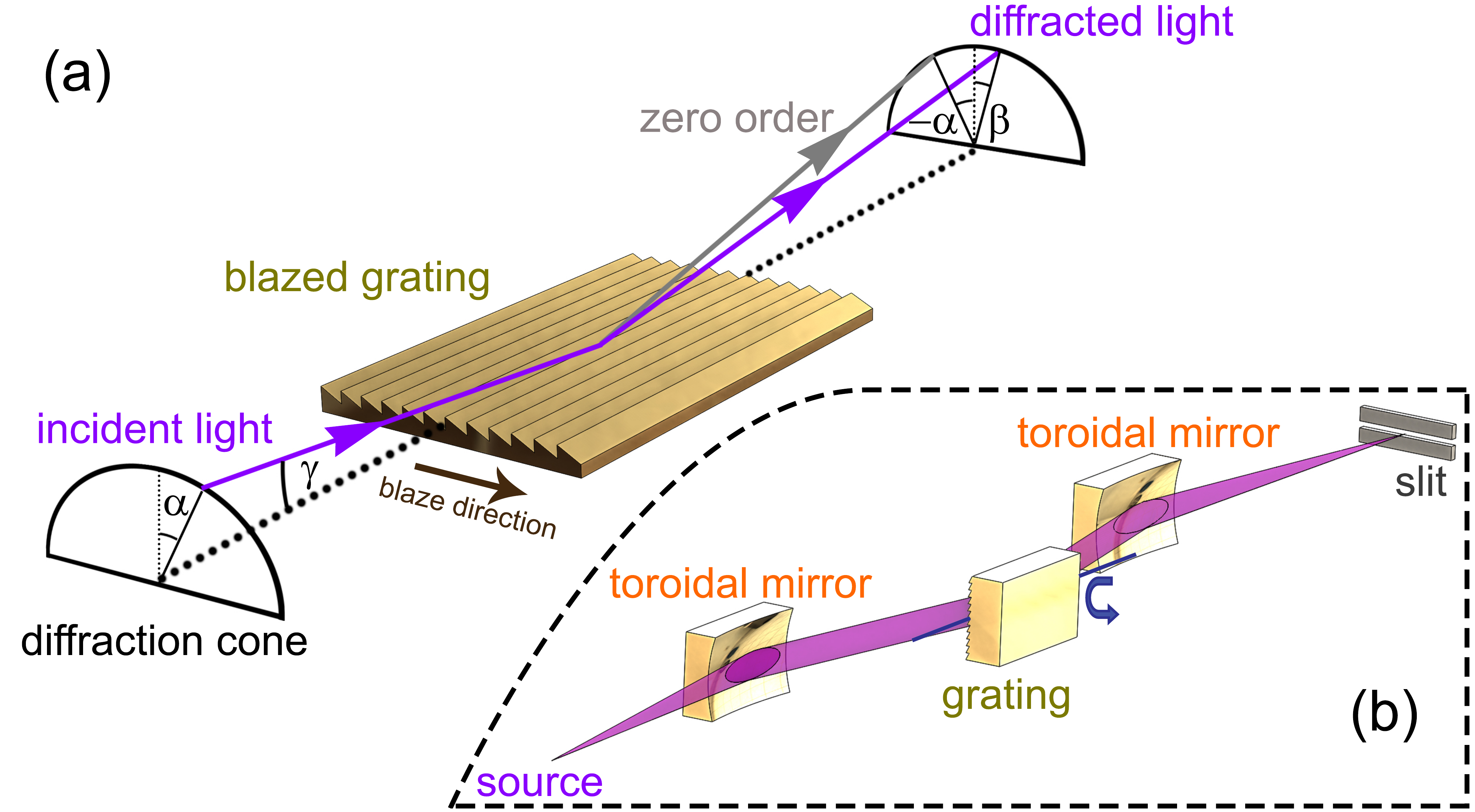}
  \caption{(a) Schematic diagram of the diffraction geometry in the off-plane mount. (b) Layout of a single Czerny-Turner monochromator stage using the off-plane configuration. The grating is rotated around an axis parallel to the grooves (indicated with a blue line) to select different photon energies.}
  \label{Fig_mono_basics}
\end{figure*}
The XUV domain puts many obstacles in the way of utilizing conventional optical designs due to the generally low reflectivity, high absorption and strong scattering of materials. One simple solution for the selection of an individual harmonic peak is using multilayer mirrors constructed from multiple alternating dielectric layers deposited on a smooth substrate surface. On the one hand, multilayer mirrors are advantageous in the preservation of pulse duration for close to normal incidence reflections, they provide high efficiency, low aberrations, and are able to achieve tight focusing conditions \cite{Wonisch2006,Hofstetter2011}. On the other hand, the lack of flexibility (a new multilayer design is desired for each harmonic frequency) and poor contrast between neighboring harmonics called for the realization of grating monochromators that use dispersive elements for spectral selection \cite{Poletto2018,Fabris2019}. The~diffraction grating can be oriented with the rulings perpendicular or parallel to the plane of incidence, depending on the favored properties of the monochromatized XUV radiation. The first, i.e. classical diffraction mount is usually preferred for providing high angular dispersion, and therefore better spectral resolution. In contrast, the design with parallel rulings, called conical diffraction mount, or off-plane mount is favored for its better temporal response and higher efficiency \cite{Poletto2006}. The grating equation for the off-plane mount configuration is given as:
  \begin{equation}
      \mathrm{sin}\gamma(\mathrm{sin}\alpha+\mathrm{sin}\beta)=m\lambda\sigma\quad,
  \end{equation} \label{eq_OPM_grating}
  where $\alpha$ and $\beta$ are the azimuths of the incident and diffracted light rays at wavelength $\lambda$ and order $m$, respectively, $\gamma$ is the altitude angle and $\sigma$ is the groove density (Fig.~\ref{Fig_mono_basics}~(a)). Although the optical design of a monochromator is usually optimized for one or the other grating orientations determined by the target application of the system, double-configuration grating monochromators are also available to provide either ultrafast time response with low spectral resolution or a longer temporal output with higher resolution in a selectable arrangement \cite{Poletto2014}. A complete XUV monochromator stage is most commonly realized in the Czerny-Turner configuration consisting of a collimating mirror, a plane grating and a focusing mirror, all in grazing incidence (Fig.~\ref{Fig_mono_basics}~(b)), although a simpler, but more expensive design utilizing a single active deformable mirror to fine tune the focusing conditions has also been reported \cite{Frassetto2013}.\par
  The use of a single diffractive element introduces variation in the optical path of an ultrashort pulse across its beam profile, a.k.a. pulse front tilt \cite{Hebling1996}, thereby leading to temporal stretching. The total time difference across the dispersed spot for wavelength $\lambda$ diffracted at order $m$ is calculated as $Nm\lambda$, where $N$ is the total number of illuminated grooves. It is possible to restore the tilted pulse front by adding a second diffraction element in a subtractive configuration resulting in a time-delay compensating monochromator \cite{Poletto2009}. More generally, by adjusting the optical path between the composing diffraction stages, customizable grating-based pulse shapers can be constructed in the XUV spectral domain for the fine compensation of the intrinsic chirp of the high harmonic radiation \cite{Mero2011}. Using a time-delay compensated arrangement, Lucchini et al. demonstrated the possibility to generate and characterize ultrashort HHG-based monochromatic XUV pulses down to 5~fs temporal duration \cite{Lucchini2018}.\par
The XUV monochromator of the \textsc{\textit{HR GHHG Condensed}} beamline is composed of two optical stages, installed in CH-06 and CH-02, respectively, containing altogether four toroidal mirrors and two plane gratings as shown in Fig.~\ref{Fig_HR-GHHG_Cond}. The monochromator is operated without an entrance slit using the HHG point as the image source. The first toroidal mirror (CTM1) collimates the light coming from the source point for the first plane grating (G1), then the second toroidal mirror (FTM1) focuses the diffracted light on the exit slit, where the light is monochromatized with a tilted pulse front. The second section (CTM2~+~G2~+~FTM2) compensates for this pulse front tilt. In order to maximize the throughput of the monochromator, all composing optics are operated at grazing incidence with the off-plane mount configurations of the gratings. Wavelength scanning is achieved by rotating the gratings in a coherent manner (one clockwise and the other counterclockwise) around the axes passing through the center of the gratings and parallel to the groove direction (see Fig.~\ref{Fig_mono_basics}~(b)).\par
The input/exit arms of the first monochromator stage were designed to be 2~m long, resulting in the total stage length of 4.5~m. Such a long first stage allows the filtering of the XUV from the residual IR via the spatial separation of the XUV and IR beams due to their differing divergences \cite{Ye2022}. The annularly shaped generating beam is removed from the XUV optical path with a holey mirror that sends the light onto a water-cooled beam dump before the first monochromator stage in CH-06 (see~Fig.~\ref{Fig_HR-GHHG_Cond}). Existing time-delay compensating monochromators have two equally constructed stages with two identical gratings and correspondingly equal input (exit) arm lengths \cite{Poletto2018b}. The use of the aforementioned annular geometry specifically designed for the high average power GHHG beamline, and the limited available laboratory space have motivated the construction of a novel, asymmetric monochromator. This geometry incorporates a long input arm of the first monochromator stage, while ensuring tight focusing conditions before the target region. The monochromator design described in this paper adopted input/output arms that are twice as long in the first section than in the second one. Accordingly, the groove densities of the G2 gratings are also doubled compared to G1 in order to zero out the pulse front tilt.\par
\vspace{\baselineskip}
  \begin{figure*}[ht]
  \centering
  \includegraphics[width=1.0\linewidth]{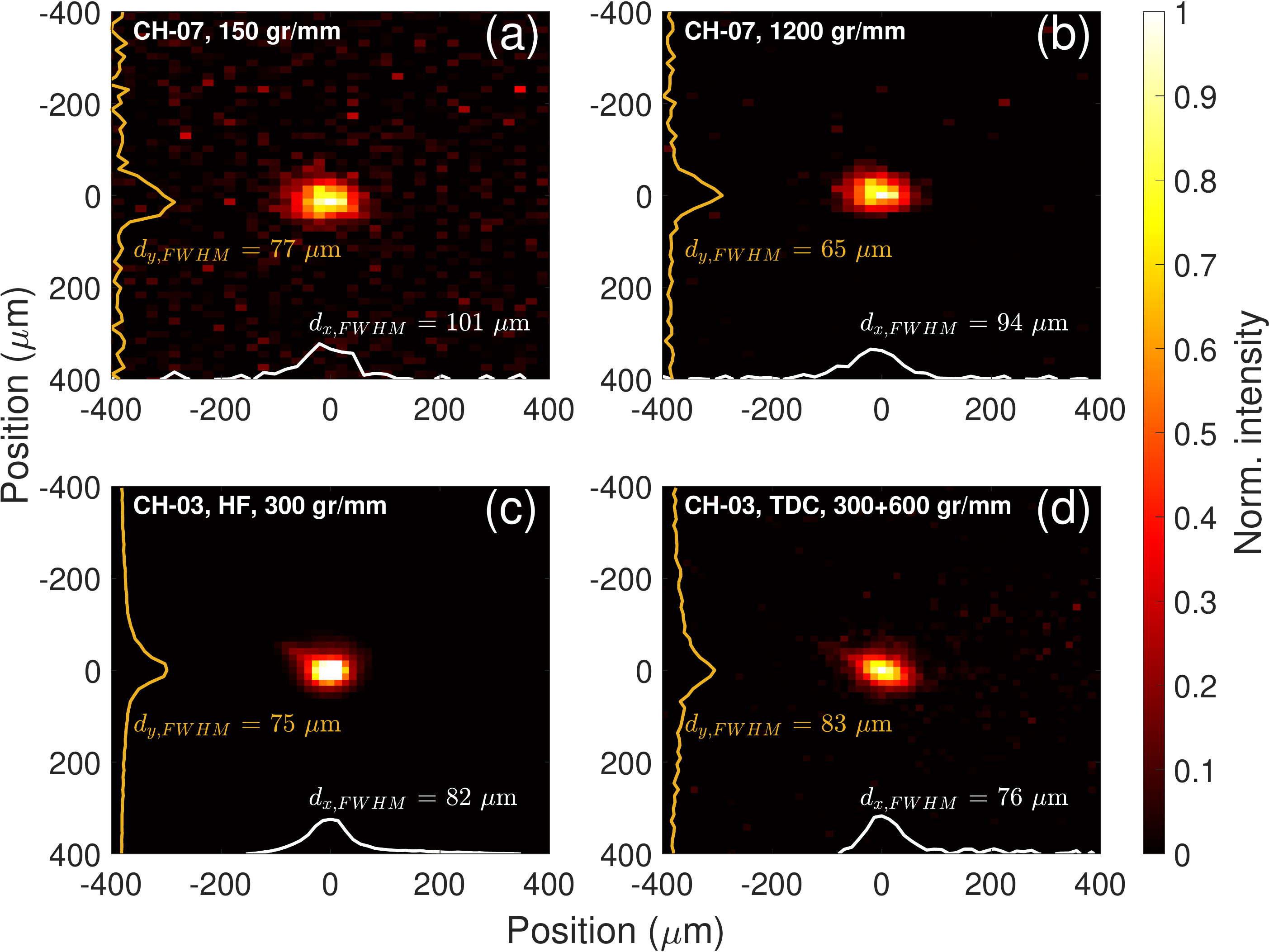}
  \caption{XUV focal spots diffracted in the zero order after the first (a,~b) and the second (c,~d) stage of the monochromator using gratings with different groove densities indicated in the top-left corner. After the second stage, the focal spot was measured both in the high-flux (HF) (c), and in the time-delay compensated (TDC) (d) operational modes.}
  \label{Fig_spatial}
\end{figure*}
  \begin{figure}[htbp]
  \centering
  \includegraphics[width=1.0\linewidth]{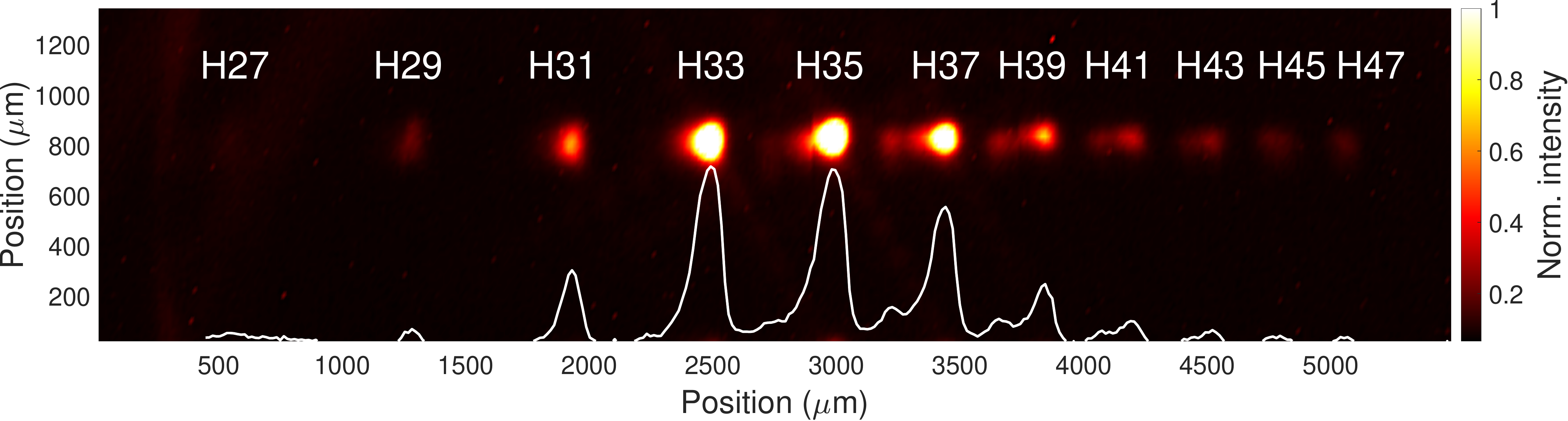}
  \caption{Raw camera image showing harmonics (H) diffracted in the first order and focused on the scintillator crystal after the first monochromator stage using the 150~gr/mm grating. The zero order condition of the same monochromator configuration is depicted in Fig.~\ref{Fig_spatial}~(a). The white curve represents the vertically integrated intensity of the beam profile.}
  \label{Fig_spatial_harmonics}
\end{figure}
The beamline can be operated in three configurations.
\begin{itemize}
    \item \textit{Broadband operation:} \par
The generated bandwidth of the high harmonic spectrum is entirely transmitted through the monochromator system.  The XUV light is reflected by two gold-coated toroidal relay mirrors (BTM1 and BTM2 in Fig.~\ref{Fig_HR-GHHG_Cond}) used in 1:1 configuration giving a total reflectivity above 80\% for s-polarized light in the operational spectral regime of the beamline (17--90~eV). The metallic filter is inserted through the filter wheel (FW in Fig.~\ref{Fig_HR-GHHG_Cond}) to filter out any residual IR contamination and to compensate for the attochirp. In the broadband operational mode, the slit (S in Fig.~\ref{Fig_HR-GHHG_Cond}) is open to avoid beam clipping.
    \item \textit{High-flux monochromatized operation:}\par
The monochromator optics of the first stage (CTM1, G1, and FTM1) are inserted, while the second stage is left in the broadband option (BTM2 inserted). The variable width slit (S) is used for the spectral selection of the diffracted radiation. Having a single diffracting section in the optical path, the photon flux of the monochromatic XUV light is maximized at the expense of the temporal response.
    \item \textit{Time-delay compensated monochromatized operation:}\par
Both of the monochromator stages are utilized, giving pulse front tilt compensated monochromatic XUV radiation with pulse durations close to the Fourier limit.
\end{itemize}
The monochromator accommodates different gratings, which are mounted on linear translators and can be automatically selected. The main specifications of all monochromator gratings are listed in Table \ref{tab_parameters}. Three gratings are used for high spectral resolution (50--800~meV), and two for low resolution (100--2000~meV), while providing a better temporal response in the time-delay compensated mode at the same slit width. The free spectral range and efficiency of the gratings are adjusted to cover the design bandwidth of the beamline (17--90~eV) in both modes.
\section{Diagnosis of the XUV beam}\label{sect_diagnostics}
\subsection{Focusability}
  The focusability of the instrument was tested using the zero order diffraction of the gratings both after the first (Fig.~\ref{Fig_spatial}~(a, b)), and the second stages of the monochromator (Fig.~\ref{Fig_spatial}~(c, d)) by recording, with CMOS cameras, fluorescent light from the Ce:YAG scintillator crystals installed after each stage. The crystals were placed at the position of the slit (CH-07) and the target region of the TOF spectrometer (CH-03), respectively, where the two after-stage focii are located. Figure~\ref{Fig_spatial} demonstrates nice, aberration-free (down to the source size) XUV spots having full width at half maximum (FWHM) diameters below $100~\upmu$m after both stages. The slit was removed during the measurements. Note~that in CH-07 the Ce:YAG crystal---irradiated normal to the surface---was tilted by 45 degrees with respect to the camera around a vertical axis in the laboratory frame, while the crystal in CH-03 was observed normally to the surface, but was tilted by 45 degrees both around the beam propagation axis and around the global vertical axis. Such geometries were taken into account in the visualization of the recorded spots in Fig.~\ref{Fig_spatial} and in the calculation of the FWHM beam diameters of the vertical and horizontal integrated beam profiles depicted in each subset by the orange and white curves, respectively. The spot sizes were found to be between 70 and 100 $\upmu$m in case of all operational modes of the monochromator as demonstrated in Fig.~\ref{Fig_spatial}: for two gratings with the lowest (a) and highest (b) groove densities, as well as for the high-flux (c) and time-delay compensated (d) operational modes. The spot sizes in the diffracted beam were also measured and they were found to be almost constant for the different harmonics and similar to the zero order spot size. Figure~\ref{Fig_spatial_harmonics} shows the harmonic spots in the first order diffraction with the image quality optimized for the 35$^\mathrm{th}$ harmonic order (propagating through the optical axis of the focusing toroidal mirror at this particular azimuth angle of the grating). Although the focusing conditions for different harmonics can be different due to the inherent wavelength dependence of HHG \cite{Quintard2019,Wikmark2019,Hoflund2021,Veyrinas2021}, the measured diameters for a given harmonic were found to be similar after each monochromator stage and larger than the diffraction-limited size indicating that they are limited by the finite spot size of the XUV source (imaged in a one-to-one ratio throughout the beamline), which is estimated to be around 60~$\upmu$m, assuming a $1/\sqrt{6}$ scaling with the estimated laser spot size of 145~µm \cite{Hadrich2010}.
  \subsection{Spectral bandwidth}
\begin{figure*}[ht]
  \centering
  \includegraphics[width=1.0\linewidth]{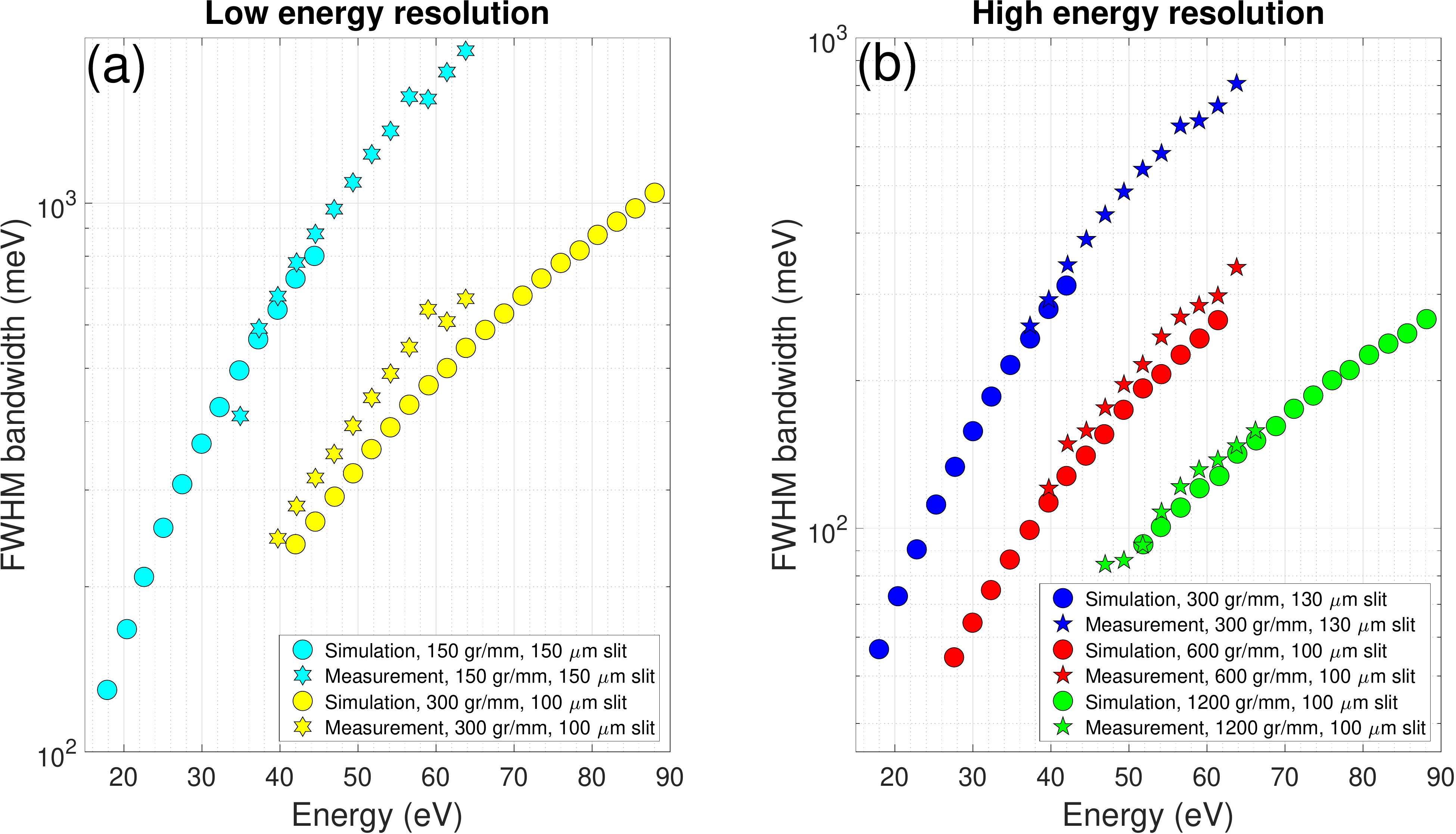}
  \caption{FWHM bandwidth of the selected high harmonic radiation in the low (a) and high (b) energy resolution modes (circle – ray-tracing simulation, star – measurement).}
  \label{Fig_spectral}
\end{figure*}

  Prior to testing the spectral and temporal performances of the XUV monochromator, the system was spectrally calibrated using the following procedure:
  \begin{enumerate}[label=(\roman*)]
  \item The first stage was set to zero order, and the position of the focused beam on the Ce:YAG crystal in CH-07 was recorded. By acting on the pitch of the plane grating (i.e. the rotation around an axis parallel to the grating surface and to the grooves, illustrated by a blue line in Fig.~\ref{Fig_mono_basics}~(b)) with a calibrated stepper motor, it was assured that the focal spot of the zero order beam is centered on the slit. The pitch value in degrees corresponding to this situation was handled as an offset during calibration.
  \item The pitch of the grating was changed until harmonics appeared on the crystal from the first order diffraction. Each harmonic was centered on the slit and the offset-corrected pitch values ($\Delta\beta_{\mathrm{measured}}=\beta+\alpha$, see Fig.~\ref{Fig_mono_basics}) were recorded.
  \item By knowing the separation between consecutive harmonics (2.41~eV using a generation beam centered at 1030~nm), the azimuth of the diffracted and incident rays ($\Delta\beta_{\mathrm{theoretical}}$) were calculated for assumed harmonic wavelengths ($\lambda$) by using the grating equation for the off-plane geometry:
    \begin{equation}
      \Delta\beta_{\mathrm{theoretical}}=\mathrm{sin}^{-1}\frac{m\lambda\sigma}{2\mathrm{sin}\gamma}\quad,
  \end{equation}
  where $m$=1 is the order of diffraction, $\sigma$ and $\gamma$ are the groove density and the altitude angle, respectively, both are specified by the grating manufacturer.
  \item The assumptions about on the $\lambda$ harmonic wavelength values and the $\gamma$ altitude angle were adjusted for the best possible agreement between $\Delta\beta_{\mathrm{theoretical}}$ and $\Delta\beta_{\mathrm{measured}}$.
  \item Steps (i)-(iv) were repeated for all five gratings in the first stage of the monochromator making sure that close agreement between the used $\lambda$ and $\gamma$ values should be obtained for all gratings. Special attention was paid to guarantee the latter during the installation and manual alignment of the gratings.
  \end{enumerate}
  The spectral performance of the monochromator was validated by measuring the separation of harmonics on the Ce:YAG screen in CH-07. The results are presented in Figs. \ref{Fig_spectral} (a) and (b) in the low, and high energy resolution modes, respectively, revealing that the measured (marked with circles) and the theoretically achievable minimum (marked with stars, determined using ray-tracing simulation) spectral selectivity values are close to one another in case of all plane gratings. On the one hand, in the high resolution option a bandwidth in the order of 100 meV or even narrower is achievable for harmonics below 35 eV. On~the other hand, in the low energy resolution mode it is possible to set a bandwidth that is high enough to separate two consecutive harmonics over the whole working photon energy range of the \textsc{\textit{HR GHHG Condensed}} beamline. In this way, it is feasible to separate individual harmonics while preserving their total natural bandwidth, and thereby achieving better flux compared to a narrow bandwidth selection.
 \subsection{Time duration of pulses}
 Measuring the duration of monochromatic XUV pulses is a challenging task due to their ultrashort nature (few fs), energetic spectral range (few nm) and relatively low flux, i.e. features which prevent, in one way or another, the utilization of conventional pulse characterization methods, such as electronic sampling or XUV-XUV autocorrelation \cite{Tzallas2003,Makos2020}. Nevertheless, a cross-correlation scheme can still be implemented by combining spatially and temporally the monochromatic XUV radiation with a weak portion of the generating IR field \cite{Lucchini2018,Murari2020}. The spatial overlap was ensured using the Ce:YAG crystal in CH-03, while the temporal overlap was found first by using a combination of an ultrafast photodiode and an oscilloscope (Tektronix MSO/DPO70000) with few ps resolution, and then---more~accurately---by monitoring the IR-IR pump-probe interference pattern on a beam profiler (Thorlabs - BC106N-VIS/M) at a low nominal output power ($\approx$1~W) of the primary laser source. A motorized iris (I1) in CH-01 was used to decrease the IR intensity to the $10^{11}$ – $10^{12}$ W/cm$^2$ range in the XUV-IR interaction volume.\par
 Argon gas atoms were ionized by the XUV light in the interaction region of the TOF electron spectrometer. If the XUV photons have enough energy to overcome the ionization potential of the target atom, and in absence of particular atomic structures or resonances, the temporal properties of the XUV radiation will be directly imprinted into the generated electron burst \cite{Lucchini2018}. The photoelectron spectrum will thus resemble the photon spectrum. In the presence of an IR field, the electron momenta get modulated before detection. By scanning the relative delay between the IR and XUV pulses, we obtain a collection of modulated electron spectra, which is called spectrogram. The spectrogram of a single harmonic selected by the monochromator consists of a single emission line corresponding to direct ionization via the absorption of an XUV photon from the harmonic. Additional sideband (SB) lines of a given length spaced by $\pm\hbar\omega_{\mathrm{IR}}$ around the XUV photoelectron peak appear; these lines correspond to two-color ionization involving one XUV photon and at least one IR photon with the central angular frequency of $\omega_{\mathrm{IR}}$. In case of a single harmonic, this SB signal does not oscillate, as only one pathway contributes to the formation of a SB compared to the spectrograms obtained with a comb of harmonics \cite{Ye2020,Ye2022,Maroju2020}. Nevertheless, the SB signal is still sensitive to the XUV and IR intensity envelopes, and it can be used to estimate the time durations of the pulses \cite{Moio2021}.\par
  \begin{figure*}[ht]
  \centering
  \includegraphics[width=1.0\linewidth]{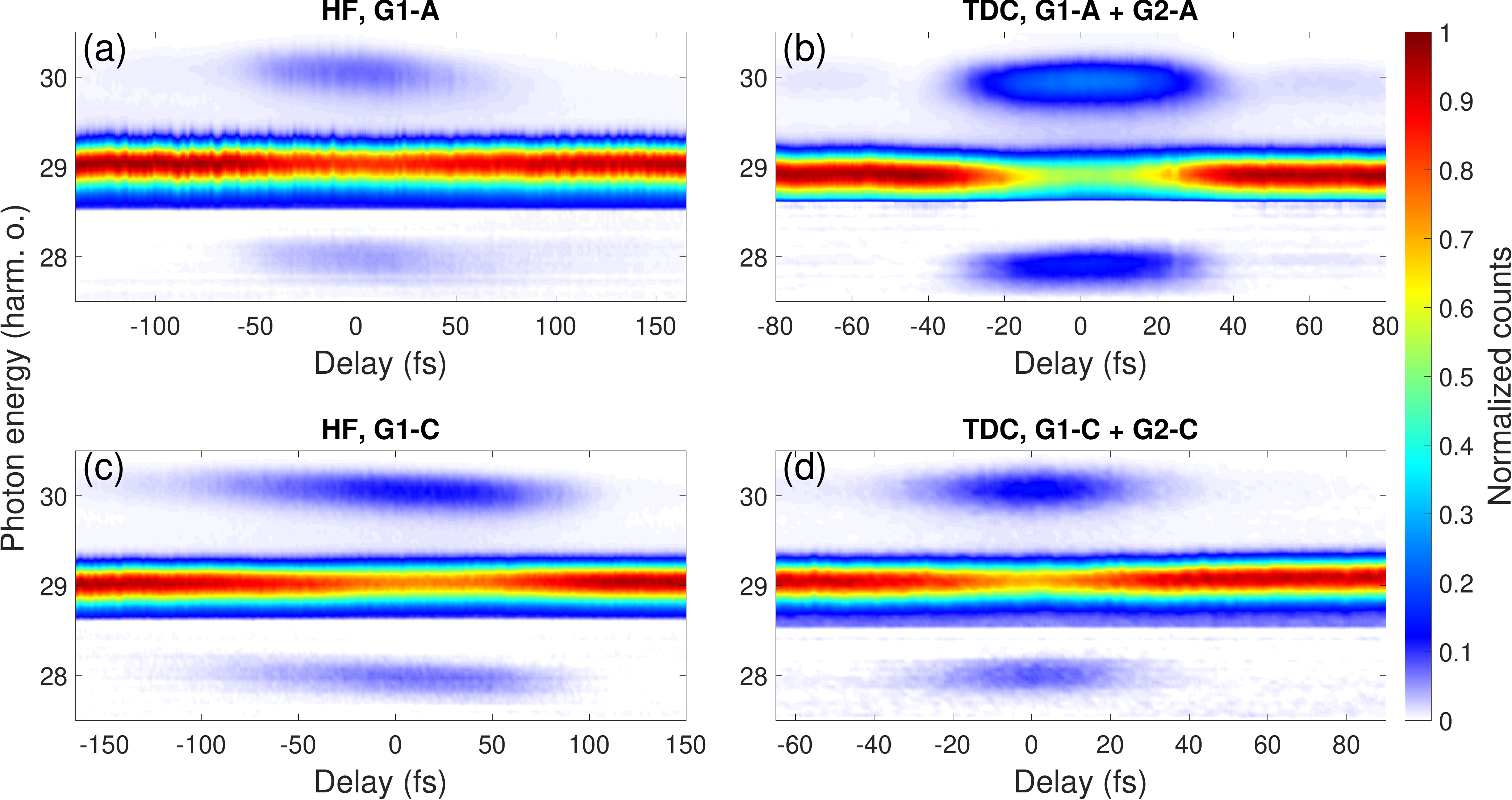}\\
  \vspace{0cm}
  \hspace*{-1cm}\includegraphics[trim=3.5cm 0cm 4.7cm 0cm,clip,width=1.0\linewidth]{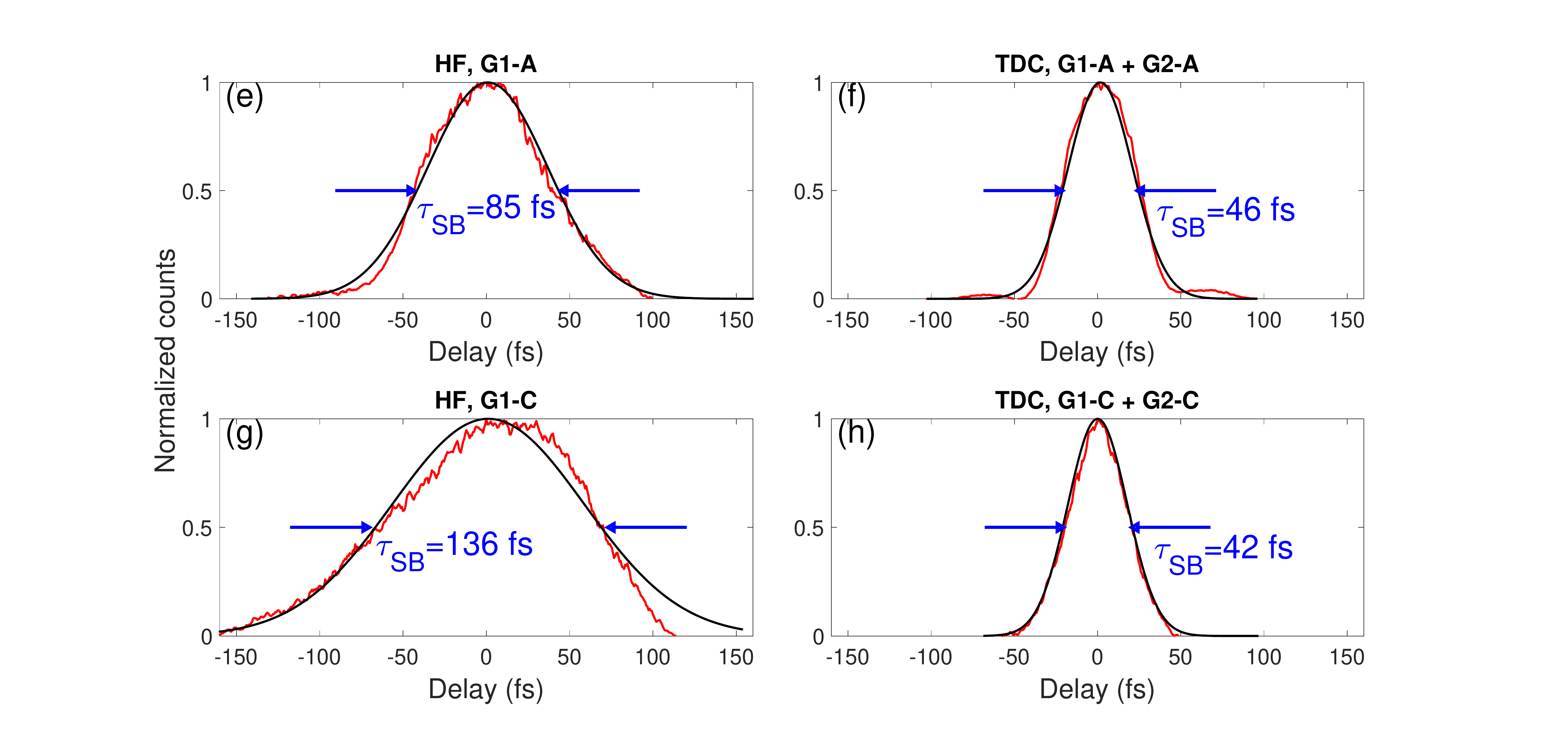}
  \caption{Experimental pump-probe traces recorded in low (a, b) and high resolution (c, d) mode with the selection of the 29$^\mathrm{th}$ harmonic without (a, c) and with (b, d) time-delay compensation using the double grating (TDC) and high-flux (HF) setups}, respectively. e--h: Time-dependent spectrally integrated signals of the upper SBs depicted in subplots a--d, respectively. The specifications of the gratings (G1-A, G2-A, G1-C and G2-C) used in the measurements are listed in Table~\ref{tab_parameters}.
  \label{Fig_temporal}
\end{figure*}
  \subsubsection{Long driving pulses}
  A first and easy approach to estimate the XUV pulse duration is by calculating the SB duration from the SB yield as a function of the XUV-IR temporal delay. If the XUV photon energy is high enough to allow the XUV field to dominate the continuum electron dynamics during photoionization, the core potential can be treated as a small perturbation ignored to the lowest order (strong field approximation, SFA \cite{Kitzler2002,Moio2021}). This results in the following expression for the photoelectron kinetic energy spectrogram $S(\boldsymbol{p},\tau)$, in atomic units:
  \begin{equation}
      S(\boldsymbol{p},\tau)=\bigg|\int_{-\infty}^{+\infty}\mathrm{d}tX_{\mathrm{si}}(t+\tau)e^{i\phi(\boldsymbol{p},t)}e^{i\big(\frac{p^2}{2}+I_\mathrm{p}\big)t}\bigg|^2\quad\mathrm{, in~which}
  \end{equation}
  \begin{equation}
      \phi(\boldsymbol{p},t)=-\int_t^\infty\mathrm{d}t'(\boldsymbol{p} \cdot \boldsymbol{A}_{\mathrm{IR}}(t')+A^2_{\mathrm{IR}}(t')/2)\quad,
  \end{equation}
  where $X_{\mathrm{si}}(t)$ is the photoelectron wavepacket, which, neglecting the resonances and strong modulations of the dipole moment of the target rare gas atom \cite{Yakovlev2010,Borrego-Varillas2021}, can be approximated with the temporal profile of the XUV field, $E_{\mathrm{XUV}}(t)$. $I_\mathrm{p}$ is the gas ionization potential, $\boldsymbol{p}$ is the final electron momentum and $\tau$ is the XUV-IR temporal delay. If the IR intensity is set to generate only the first SB order, and both the IR and the XUV pulses can be considered under the conditions of the slowly varying envelope approximation, then it is possible to show that the SB signal follows the cross-correlation between the IR and XUV pulses \cite{Lucchini2018}. Moreover, assuming that the pulses have a Gaussian temporal profile, and knowing the IR time duration ($\tau_\mathrm{IR}$), the XUV time duration ($\tau_\mathrm{XUV}$) can be obtained from the SB temporal width ($\tau_\mathrm{SB}$) with the simple relation:
  \begin{equation}\label{eq_XUV_dur_simple}
      \tau_\mathrm{XUV,meas}=\sqrt{\tau_\mathrm{SB}^2-\tau_\mathrm{IR}^2}\quad.
  \end{equation}
  To achieve such conditions, the HR-1 laser system was used with a single compression stage delivering multicycle laser pulses. The FWHM pulse duration close to the HHG point was measured to be $\tau_\mathrm{IR}$=38~fs by frequency-resolved optical gating (FROG) \cite{Gallmann2000}.
  Figure~\ref{Fig_temporal} shows the experimental photoelectron spectrograms measured at the output of the monochromator operating in the low (a, b) and high energy resolution modes (c, d) with the selection of the 29$^{\mathrm{th}}$ harmonic. For a direct demonstration of pulse front tilt compensation, the measurements conducted in high-flux mode (a, c) were repeated in the time-delay compensated configuration (b, d). Every spectrum was normalized to the total count at a fixed delay in order to decrease the effect of harmonic signal fluctuation. The measurements in the low resolution option show that SB durations, written in blue in Fig.~\ref{Fig_temporal}, of around 85~fs (e) were cut to around 46~fs (f) when the second grating stage was also used.\par
  The measurements were repeated in the high energy resolution mode, where a narrow bandwidth of 120~meV was selected by adjusting the slit width to 70~$\upmu$m. Here, uncompressed XUV pulses yielding SB durations of 136~fs (Fig.~\ref{Fig_temporal}~(g)) were shortened to only around 42~fs (Fig.~\ref{Fig_temporal}~(h)), which is similar to the duration obtained in the low energy resolution mode.\par
  The temporal broadening $\tau_p$ due to induced pulse front tilt can be described by the following equation \cite{Hebling1996}:
  \begin{equation}\label{eq_broadening}
    \tau_p=\frac{\lambda}{c}\frac{\sigma mS}{\mathrm{cos}\Delta\beta}\quad,
  \end{equation}
  where $\lambda$ is the central wavelength, $\sigma$ is the groove density, $m$ is the diffraction order, $S$ is the focused XUV spot size, $c$ is the speed of light, and $\Delta\beta$ is the azimuth angle.\par
  According to Eq.~\ref{eq_broadening}, a grating pair with double groove density in the setup (which is the case for the grating pairs tested in the high energy resolution mode, see Table \ref{tab_parameters}) should result in a pulse front tilt twice as large. The measured shortenings of around 40 and 95~fs in the low and high energy resolution options, respectively, are in good agreement with the groove densities of the grating pairs used in these two monochromator configurations.\par
In order to further benchmark the reconstruction procedure, the measurements were repeated at a different XUV photon energy (corresponding to the 35$^\mathrm{th}$ harmonic) and a rough estimation on the temporal response of the monochromator was also performed (Table \ref{tab_pulse_dur}). Here the transform-limited XUV pulse durations were calculated either from the adjusted spectral selection of the monochromator (in the high energy resolution mode) or from the XUV photon spectrum recorded with the flat-field XUV spectrometer (in the low energy resolution mode), see the third and fourth columns in Table \ref{tab_pulse_dur}.\par
Two effects were considered on the ultrafast pulse that lead to pulse elongation even in the time-delay compensated configuration. The first one was the group delay dispersion (GDD) introduced in the time-delay compensated operation (fifth column in Table \ref{tab_pulse_dur}). Analogously to grating compressors for the visible-infrared spectral range \cite{Treacy1969,Strickland1985}, the time-delay compensated operational mode of the monochromator is considered as an XUV pulse shaper that introduces a controllable GDD. The optical path decreases linearly with the wavelength, and this forces the GDD to be almost constant and positive. In particular, its value depends on the chosen grating, the photon energy and the actual XUV bandwidth \cite{Frassetto2008,Mero2011}. For the current work, the residual GDD was estimated with a custom ray-tracing program \cite{Frassetto2008}, with which the optical path lengths of the rays propagating through the monochromator ($l(\omega_{\mathrm{XUV}})$) could be calculated and the GDD could be derived using the formula:
\begin{equation}
    \mathrm{GDD}(\omega_{\mathrm{XUV}})=\frac{1}{c}\frac{\partial l(\omega_{\mathrm{XUV}})}{\partial\omega_{\mathrm{XUV}}}\quad.
\end{equation}
For the 29$^{\mathrm{th}}$ and 35$^{\mathrm{th}}$ harmonics selected in the low energy resolution mode, the residual GDD was estimated to be 12 fs$^2$ and 4 fs$^2$, respectively. Correspondingly, for the same harmonics selected in the high energy resolution scheme a GDD of 40 fs$^2$ and 20 fs$^2$ is expected for bandwidths of 120 and 210~meV. The second effect was the compensation of the pulse front tilt due to diffraction. This is accomplished when all the rays that have equal wavelength and are emitted in different directions by the high harmonic source travel the same optical path. Ideally, the compensation is perfect for a double-grating configuration, although a slight misalignment in the optical path and/or distortions introduced by the imaging optics may give some residual aberrations of the pulse front, which was estimated to be below 10~fs by ray tracing simulations (column six).
Both methods give coherent results for the pulse durations of the studied monochromatic XUV pulses between 13 and 27~fs (corresponding to the uncertainty of the simple reconstruction based on Eq.~\ref{eq_XUV_dur_simple}) after time-delay compensation. As a rule of thumb, the duration of a single harmonic pulse should roughly equal half the duration of the driving IR pulse due to transient phase matching at a typical few percent ionization rate in case of the applied generation conditions \cite{Christov1996,Lucchini2018}. Therefore, these values are consistent with the relatively long ($\tau_\mathrm{IR}\approx$40~fs) generating fundamental field used during the XUV-IR cross-correlation measurements.
  \begin{table}[ht]
    \centering
    \resizebox{\columnwidth}{!}{\begin{tabular}{|m{23.5mm}|m{9mm}|m{15mm}|m{12mm}|m{7.5mm}|m{16.2mm}|m{11mm}|m{13.4mm}|}
    \hline
    \rowcolor[gray]{.9}[0.8\tabcolsep]
    Mode & Harm.\newline order & Bandwidth (meV) & FL pulse\newline duration \newline(fs) & GDD \newline(fs$^2$) & Aberrations \newline(fs) & $\tau_{\mathrm{XUV,est}}$ \newline(fs) & $\tau_{\mathrm{XUV,meas}}$ \newline(fs)\\ [1ex]
    \hline\hline
    \multirow{2}{*}{Low energy res.} & \multicolumn{1}{c|}{29} & \multicolumn{1}{c|}{400} & \multicolumn{1}{c|}{4.6} & \multicolumn{1}{c|}{12} & \multicolumn{1}{c|}{<10} & \multicolumn{1}{c|}{<$\mathbf{18.6}$}& \multicolumn{1}{c|}{$\mathbf{22.7}$}\\\cline{2-8}
    & \multicolumn{1}{c|}{35} & \multicolumn{1}{c|}{400} & \multicolumn{1}{c|}{4.6} & \multicolumn{1}{c|}{4} & \multicolumn{1}{c|}{<10} & \multicolumn{1}{c|}{<$\mathbf{15.2}$}& \multicolumn{1}{c|}{$\mathbf{15.8}$}\\\cline{1-8} 
    \hline\hline
    \multirow{2}{*}{High energy res.} & \multicolumn{1}{c|}{29} & \multicolumn{1}{c|}{120} & \multicolumn{1}{c|}{15.2} & \multicolumn{1}{c|}{40} & \multicolumn{1}{c|}{<10} & \multicolumn{1}{c|}{<$\mathbf{26.8}$}& \multicolumn{1}{c|}{$\mathbf{12.8}$}\\\cline{2-8}
    & \multicolumn{1}{c|}{35} & \multicolumn{1}{c|}{210} & \multicolumn{1}{c|}{8.7} & \multicolumn{1}{c|}{20} & \multicolumn{1}{c|}{<10} & \multicolumn{1}{c|}{<$\mathbf{20.8}$}& \multicolumn{1}{c|}{$\mathbf{26.5}$}\\\cline{1-8}
\end{tabular}}
\caption{Comparison of estimated and measured XUV pulse durations.}
\label{tab_pulse_dur}
\end{table}
\subsubsection{Short driving pulses}
In order to assess and validate the limit of the temporal capabilities of the monochromator using spectrally broad harmonics, the laser was set up in the short pulse mode utilizing both post-compression stages and providing pulses down to 6~fs duration. The pulse duration was verified at the laser output using second-harmonic dispersion scan \cite{Miranda2012}. In the generation chamber of the \textsc{\textit{HR GHHG Condensed}} beamline, the thickness of two fused silica wedge pairs (W1 and W2 in Fig.~\ref{Fig_HR-GHHG_Cond}) were increased finely and simultaneously to introduce dispersion until the harmonic peaks in the XUV spectrum became distinguishable, but not fully separated, stretching the laser pulses to approximately 10--15~fs. At around 34.9~eV, a single peak (corresponding to the 29$^\mathrm{th}$ harmonic) was selected with an FWHM bandwidth of 700~meV, and XUV-IR cross-correlation traces were recorded by ionizing Ar atoms in the TOF electron spectrometer.\par
The Frequency-Resolved Optical Gating for Complete Reconstruction of Attosecond Bursts (FROG CRAB) technique was developed for the retrieval of the amplitude and phase of attosecond XUV fields from two-color cross-correlation spectrograms \cite{Mairesse2005}. However, the selection of a narrow bandwidth from the broadband spectrum results in the loss of sub-cycle resolution in the FROG CRAB trace. This reduced information hinders the application of the most commonly used reconstruction techniques in the case of single harmonic few-femtosecond XUV pulses. For this reason, we have used the combination of FROG CRAB with the extended ptychographic engine (ePIE), which has already proven its competence to temporally characterize ultrashort XUV pulses produced by HHG and spectrally selected by a monochromator \cite{Lucchini2018}. Moreover, it was demonstrated that ePIE has a variety of advantageous traits in comparison with other iterative pulse reconstruction techniques \cite{Gallmann2000,Chini2010}, such as outstanding convergence, robustness to white noise and capability to work with non-equidistant sampling of the delay axis \cite{Spangenberg2015,Lucchini2015_ptycho}.\par
\begin{figure*}[ht]
  \centering
  \includegraphics[trim=0.75cm 0cm 1.25cm 0cm,clip,width=1.0\linewidth]{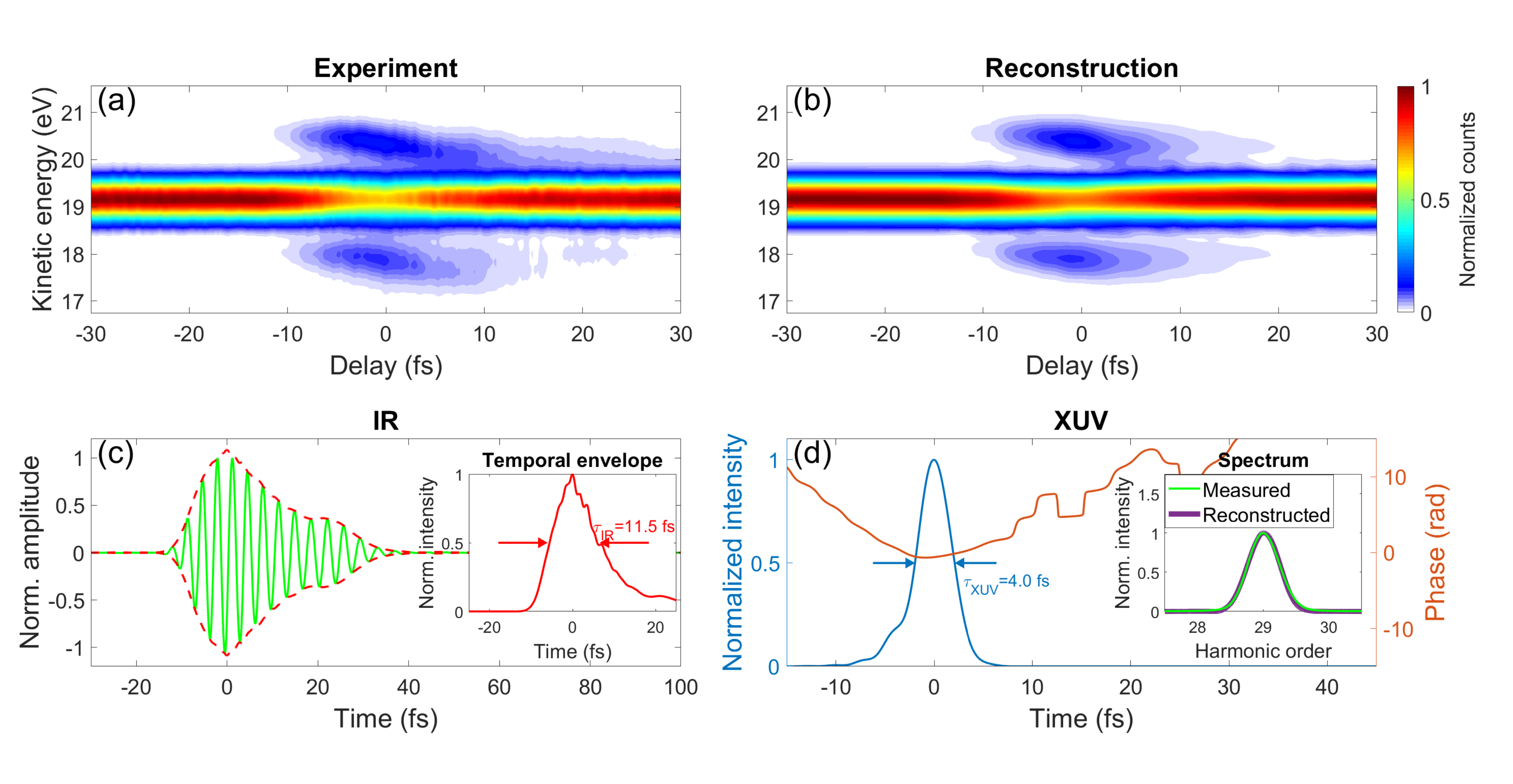}
  \caption{Complete temporal characterization of monochromatic XUV pulses using the ptychographic reconstruction technique. (a) Experimentally recorded and (b) retrieved spectrograms. (c) The reconstructed temporal profile of the electric field amplitude (solid green) and envelope (dashed red). The temporal intensity envelope (solid red) of the same pulse is represented in the small inset. Subplot (d) displays the reconstructed XUV temporal intensity (light blue) and phase (brown). The inset shows the comparison of the retrieved XUV spectrum (purple) to the one measured using an XUV photon spectrometer (green). The spectral bandwidth was found to be 700~meV (FWHM) corresponding to the transform-limited pulse duration of 3.0~fs. The initial conditions of this particular reconstruction were: 7 and 4 fs IR and XUV transform-limited pulse durations, respectively, and 25 fs$^2$ IR chirp. Using a wide set of initial values, the temporal durations were retrieved as 12.1±0.6 fs, and 4.0±0.2 fs for the IR and XUV pulses, respectively.}
  \label{Fig_ePIE}
\end{figure*}

Figure~\ref{Fig_ePIE}~(a) and (b) present the experimentally recorded spectrogram and its ePIE reconstruction, respectively. A quasi-parallel tilt in both the bottom and top SBs is clearly visible in both traces, indicating the presence of temporal chirp in the monochromatic XUV pulses \cite{Moio2021}. As initial conditions, the Fourier-limited pulse durations of the IR and XUV pulses were assumed to be 7 and 4~fs, respectively, and the chirp of the IR field was taken as 25~fs$^2$. The ePIE algorithm was iterated 1000 times until good agreement was obtained between the reconstructed and experimental traces. The reconstruction of the temporal profile of the IR and XUV fields are shown in Fig.~\ref{Fig_ePIE}~(c) and (d), respectively. The validity of the reconstruction was cross-checked by comparing the reconstructed XUV spectrum (marked with a solid purple line in the small inset of Fig.~\ref{Fig_ePIE}~(d)) to a spectrum recorded by the flat-field XUV spectrometer (solid green line in the same subplot). The good agreement between the two spectra further supports the correct convergence of the algorithm. The reconstruction was repeated several times with random initial guesses for the transform-limited XUV and IR pulse durations, as well as for the chirp of the driving field in order to validate the correct divergence and obtain the error of the reconstruction. In this way, the temporal durations of the IR and XUV pulses were determined to be 12.1$\pm$0.6~fs, and 4.0$\pm$0.2~fs, respectively. The latter value is slightly smaller than the shortest XUV pulse duration (5.0$\pm$0.5~fs) reported so far at the output of a time-preserving monochromator \cite{Lucchini2018}.
\subsection{Efficiency and photon flux}
\begin{figure*}[ht]
  \centering
  \includegraphics[trim=0.9cm 0cm 3.2cm 0cm,clip,width=1.0\linewidth]{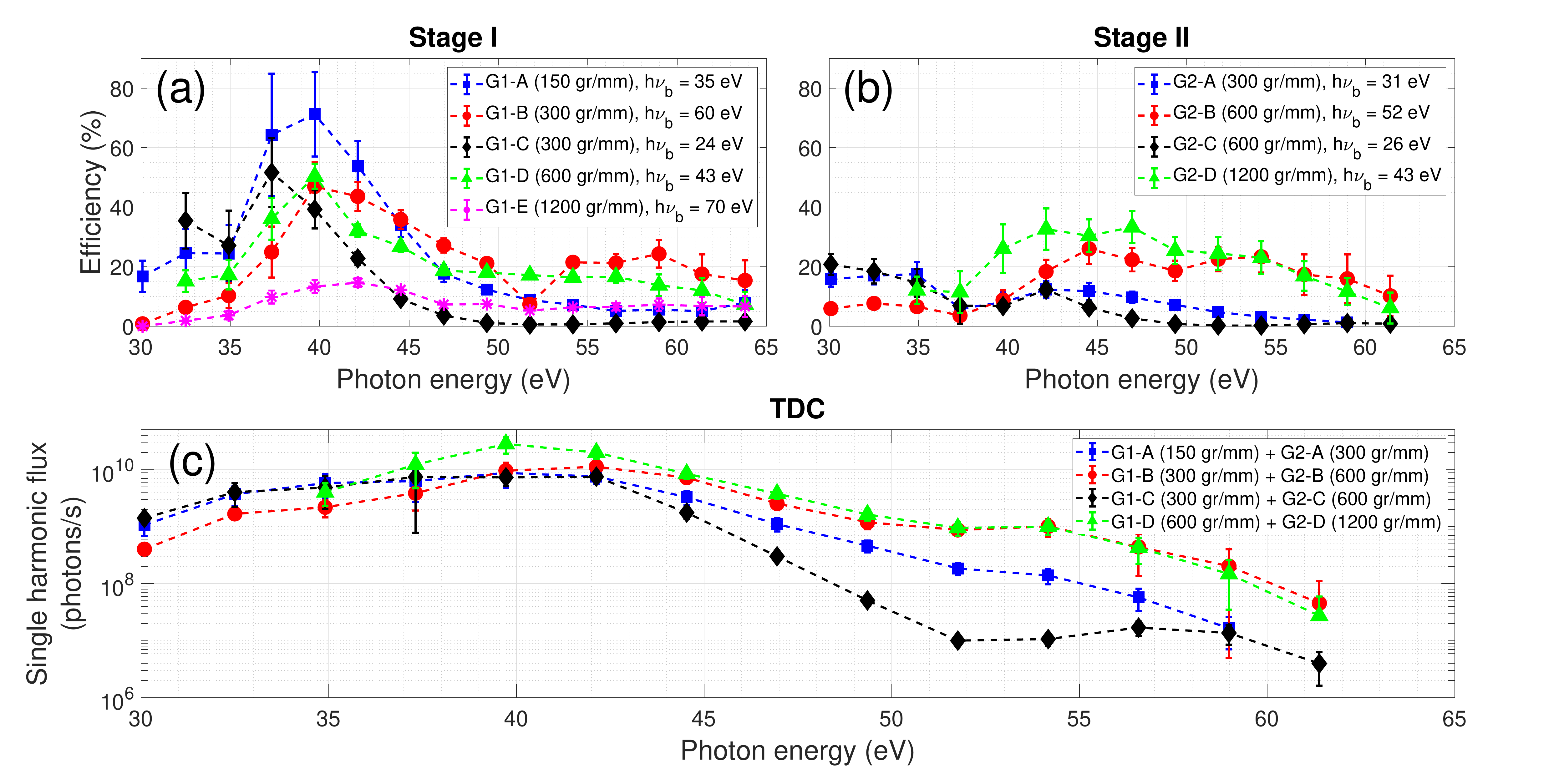}
  \caption{Efficiency of the first (a) and second (b) monochromator stages in the first diffraction order with respect to the broadband configuration. (c) Absolute XUV photon flux of a single harmonic measured at the output of the monochromator in the time-delay compensated operational mode. The error bar lengths correspond to the standard deviation in the measured data.}
  \label{Fig_eff}
\end{figure*}

The overall transport efficiency of the monochromator and the output photon flux are critical parameters for checking the feasibility of an experiment and determining how long that experiment will take to produce reasonable statistics. In order to determine the efficiency of the first (second) monochromator stage, the grating in the first (second) stage was operated in the first order diffraction, while only a single toroidal mirror was left in the beam path in the remaining stage. The slit width was set to the maximum (450~$\upmu$m) to allow the largest possible bandwidth of a single harmonic to be transmitted through the XUV monochromator. The spectra of the produced radiation were recorded with the flat-field XUV spectrometer located in CH-05 (see Fig.~\ref{Fig_HR-GHHG_Cond}) and the signal integrated in a single harmonic peak was compared to that of in the broadband operation. Figure~\ref{Fig_eff}~(a) and (b) show the efficiencies of the first, and second monochromator stages, respectively. For each grating, the blaze photon energy (h$\nu_\mathrm{b}$) is indicated in the figure legend. Figure~\ref{Fig_eff} reveals that, for most cases, the efficiency is higher close to the blaze conditions of the gratings, and lower for gratings with higher groove densities, when the amount of radiation diffracted in the zero order increases. Due to the off-plane mount of the gratings and all optics operating at grazing incidence, a relatively high efficiency can be reached, i.e. 70$\pm$14\% for the first (using G1-A), and 33$\pm$11\% for the second stage (using G2-D) at photon energies of 40 and 47~eV, respectively. The efficiency in the time-delay compensated operational mode can be calculated by multiplying the efficiencies of the two conjoined monochromator stages.\par
Finally, the total output flux in broadband operation was measured by an XUV photodiode inserted into the beam path after the second monochromator stage (see Fig.~\ref{Fig_HR-GHHG_Cond}). The photocurrent measured on the PD output was converted to XUV pulse energy using the calibration table from NIST. In this way, the broadband XUV pulse energy was determined to be 76$\pm$3~pJ per pulse, corresponding to the XUV flux of 1.27$\pm0.05\times10^{12}$~photons/s in a spectral range between 22.9 and 63.8~eV. Figure~\ref{Fig_eff}~(c) shows the measured absolute photon flux of single harmonics selected in the time-delay compensated mode of the monochromator. The maximum photon flux obtained was 2.8$\pm0.9\times10^{10}$~photons/s at 39.7~eV (33$^{\mathrm{rd}}$~harmonic order), which, to the best of our knowledge, is the highest flux of a single harmonic reported at the output of a double stage monochromator (Table~\ref{tab_photon_flux}), and is comparable to that from small-scale synchrotron facilities \cite{Igarashi2012,Hwang2020,Chernenko2021}. Given a spot size of 0.006~mm$^2$, the repetition rate of 100~kHz and the pulse duration of 4.0$\pm$0.2~fs, this photon flux translates to 3.0$\pm1.0\times$10$^{-8}$~J/cm${^2}$ monochromatic XUV fluence, and 7.5$\pm2.5\times$10$^{6}$~W/cm${^2}$ XUV intensity per laser shot in the focal plane of the target region.
  \begin{table}[ht]
    \centering
    \resizebox{\columnwidth}{!}{\begin{tabular}{|m{25mm}|m{25mm}|m{25mm}|m{25mm}|m{25mm}|}
    \hline
    \rowcolor[gray]{.9}[1.0\tabcolsep]
   \multicolumn{1}{|c|}{Reference} & \multicolumn{1}{c|}{Photon energy (eV)} & \multicolumn{1}{c|}{Pulse duration (fs)}&\multicolumn{1}{c|}{Repetition rate (kHz)}&\multicolumn{1}{c|}{Flux (photons/s)}\\ [1ex]
    \hline\hline
   \multicolumn{1}{|c|}{\textit{Poletto et al.} \cite{Poletto2009}} & \multicolumn{1}{c|}{35.6} & \multicolumn{1}{c|}{8$\pm1~\mathrm{fs}$} & \multicolumn{1}{c|}{1} &  \multicolumn{1}{c|}{1.3$\times10^{9}$}\\\hline
   \rowcolor[gray]{.95}[1.0\tabcolsep]
    \multicolumn{1}{|c|}{\textit{Dakovski et al.} \cite{Dakovski2010}} & \multicolumn{1}{c|}{20--36} & \multicolumn{1}{c|}{N/A} & \multicolumn{1}{c|}{10} &  \multicolumn{1}{c|}{1$\times10^{9}$--1$\times10^{10}$}\\\hline
   \multicolumn{1}{|c|}{\textit{Igarashi et al.} \cite{Igarashi2012}} & \multicolumn{1}{c|}{32.8} & \multicolumn{1}{c|}{11$\pm3~\mathrm{fs}$} & \multicolumn{1}{c|}{1} &  \multicolumn{1}{c|}{5.7$\times10^{9}$}\\\hline
   \rowcolor[gray]{.95}[1.0\tabcolsep]
   \multicolumn{1}{|c|}{\textit{Yong et al.} \cite{YongNiu2017}} & \multicolumn{1}{c|}{35.7} & \multicolumn{1}{c|}{$\sim100~\mathrm{fs}$} & \multicolumn{1}{c|}{1} &  \multicolumn{1}{c|}{1$\times10^{9}$}\\\hline
   \multicolumn{1}{|c|}{\textit{von Conta et al.} \cite{vonConta2016}} & \multicolumn{1}{c|}{29.4} & \multicolumn{1}{c|}{29$\pm2~\mathrm{fs}$} & \multicolumn{1}{c|}{1} &  \multicolumn{1}{c|}{9$\times10^{8}$}\\\hline
   \rowcolor[gray]{.95}[0.95\tabcolsep]
   \multicolumn{1}{|c|}{\textit{This work}} & \multicolumn{1}{c|}{35--40} & \multicolumn{1}{c|}{4.0$\pm0.2~\mathrm{fs}$} & \multicolumn{1}{c|}{100} & \multicolumn{1}{c|}{2.8$\pm0.9\times10^{10}$}\\
  \hline
\end{tabular}}
\caption{Reported photon fluxes for the full bandwidth of a single harmonic selected by a time-preserving grating monochromator.}
\label{tab_photon_flux}
\end{table}
\section{Conclusion and outlook}\label{sect_conclusion}
We have presented the first results of a user-oriented beamline (\textsc{\textit{HR GHHG Condensed}}) equipped with a time-preserving, asymmetric XUV monochromator at the ELI ALPS user facility. The current driving laser is the HR-1 laser system at ELI ALPS. It is based on the techniques of fiber chirped-pulse amplification and multipass cell pulse compression delivering pulses down to 6~fs duration at a repetition rate of 100~kHz and with pulse energies up to 1~mJ. In the near future, the beamline will be driven by the HR-2 laser system, which is able to provide 5~mJ pulses (at the same 100~kHz repetition rate) with otherwise identical characteristics. This provides the possibility of substantially increasing the XUV flux by adjusting the focusing geometry, the gas density and the medium length via the application of pulse energy scaling concepts on the existing system \cite{Heyl2016,Heyl2016s,Weissenbilder2022}. The vacuum system, the optical components and the gas cell were constructed to be capable of accommodating the average laser output power of 0.5~kW.\par
We have also performed the detailed characterization of the generated high harmonic radiation, shaped spatially, temporally and spectrally by a two-stage time-delay compensating XUV monochromator, which was designed and implemented in a novel asymmetric geometry. Characterization measurements revealed high-flux, monochromatic XUV pulses down to 100~meV spectral bandwidth that can be focused to a minimum 70~$\upmu$m FWHM spot size in the target region. In addition, we accomplished the full (amplitude and phase) temporal characterization of monochromatic XUV pulses providing pulse durations down to 4.2±0.2 fs. Compensation of the pulse front tilt has also been evidenced by measuring significantly shorter pulse lengths in the double-grating configuration compared to the single-stage scenario.\par
For user experiments, the XUV beamline is directly connected to an energy-filtering photoemission microscope (NanoESCA), equipped with a wide range of measurement modes predestined for momentum microscopy, ARPES of very localized features, and imaging spectroscopy \cite{Escher2005,Vasilyev2016}. The beamline will be soon extended with a unique and versatile liquid jet apparatus to perform either photoelectron spectroscopy or transient absorption spectroscopy measurements. These pieces of equipment, together with the present capability of the instrument to deliver high-flux, monochromatic, tunable femtosecond XUV pulses close to their transform-limited duration, pave the way to novel time- and angle-resolved pump-probe photoemission experiments in liquid or on solid targets.

\section*{Acknowledgement}
The ELI ALPS project (GINOP-2.3.6-15-2015-00001) is supported by the European Union and co-financed by the European Regional Development Fund.

\section*{Disclosures}
The authors declare no conflicts of interest.
\bibliography{References_revised_tracked}
\end{document}